\documentclass[preprint,12pt]{aastex}
\usepackage{epsfig}
\usepackage{natbib}
\bibpunct{(}{)}{;}{a}{}{,}
\begin{document}
\def\today{\number\year\space \ifcase\month\or  January\or February\or
        March\or April\or May\or June\or July\or August\or
September\or
        October\or November\or December\fi\space \number\day}
\def\fraction#1/#2{\leavevmode\kern.1em
 \raise.5ex\hbox{\the\scriptfont0 #1}\kern-.1em
 /\kern-.15em\lower.25ex\hbox{\the\scriptfont0 #2}}
\def\spose#1{\hbox to 0pt{#1\hss}}
\def\simlt{\mathrel{\spose{\lower 3pt\hbox{$\mathchar''218$}}
     \raise 2.0pt\hbox{$\mathchar''13C$}}}
\def\simgt{\mathrel{\spose{\lower 3pt\hbox{$\mathchar''218$}}
     \raise 2.0pt\hbox{$\mathchar''13E$}}}
\def\etal{et al.\ }

\title{Stellar Evolution with Enriched Surface Convection Zones I. General
Effects of Planet Consumption}
\author{Ann Marie Cody\altaffilmark{1,2} \& Dimitar D.
Sasselov\altaffilmark{1}}
\altaffiltext{1}{Dept. of Astronomy, Harvard University, 60 Garden St.,
Cambridge MA 02138; dsasselov@cfa.harvard.edu} 
\altaffiltext{2}{Current address: California Institute of Technology, 1200 East California 
Boulevard, Pasadena, CA 91125; amc@astro.caltech.edu}

\begin{abstract}
 Abundance analyses of stars with planets have revealed
that their metallicities are enhanced relative to field stars.  Such a
trend was originally suggested to be due to accretion of iron-rich
planetary material. Based on this assumption, we have developed a stellar
evolution code to model stars with non-uniform metallicity distributions.
We have calculated ``polluted'' stellar evolution tracks for stars with
M=0.9--1.2~M$_{\odot}$. Our models encompass a range of initial metal
content from Z=0.01 to 0.03, and include metallicity enhancements within
the stellar convection zone corresponding to $\Delta$Z=0.005-0.03. We find
that the primary effects of metal enhancement on stellar structure and
evolution are expansion of the convection zone and downward shift of
effective temperature. In addition, we have computed the surface
metallicities expected for stars of different mass for fixed quantities of
pollution; there appears to be no correlation with present observational
data on the metallicities of stars known to harbor planets.
\end{abstract}

\section{Introduction}
\subsection{Metallicity Trends in Extrasolar Planet Host Stars}

Since the initial discovery of extrasolar planets in 1995
\citep{1995Natur.378..355M}, the study of these objects has grown at a rapid
rate. Thanks in part to improving Doppler technology and a number of new planet
detection techniques, more than 100 extrasolar planets are known, including
several that orbit together in systems around a common star.  But with the
wealth of data have come new puzzles, one of the most intriguing of which is an
emerging connection between stellar metallicity and the presence of planets.  
Numerous spectroscopic observations have revealed that metallicities of stars
with planets (SWPs) are enhanced on average, as compared to abundances in F-G
type field stars. Quite a few of the stars now known to be orbited by planets
have unusually high metallicities (Gonzalez, Wallerstein \& Saar
\citeyear{1999ApJ...511L.111G}; Castro \etal \citeyear{1997AJ....114..376C}).
Moreover, \citet{1998A&A...334..221G} noted that the overall metallicity {\em
distribution} is shifted relative to that of unbiased solar neighborhood star
samples. Further discussion of this offset has been offered by
\citet{2001AJ....121..432G}, and Santos, Israelian \& Mayor
\citeyearpar{2001A&A...373.1019S}. The latter find
that the mean metallicities for the two different groups are different by 0.25
dex in [Fe/H], and \citet{2002ApJ...566..442M} report similar results.  The
most recent study has been carried out by \citet{2003RvMP...75..101G} on a set
of 69 SWPs together with 43 field stars from \citet{2001A&A...373.1019S} and
\citet{2003scifro..6..392F}; they confirm the $\sim$0.2~dex significant offset
between the SWP and field star [Fe/H] distributions is seen. Statistical
analysis has enabled \citet{2001A&A...373.1019S} to determine a probability of
10$^{-7}$ that the two distributions are the same. Despite the convincing
results, there is some concern that biases in SWP detection methods are
responsible for the appearance that SWPs are more metal rich than average. It
is well known that the current Doppler searches favor identification of massive
planets in small orbits, but reasons for a possible association between
close-in orbits and high-metallicity parent stars are not clear. Nevertheless,
several analyses \citep{2002ApJ...566..442M,2003RvMP...75..101G} have addressed
these issues and concluded that the combination of potential biases still
cannot generate the observed metallicity differences. Because the number of
known extrasolar planets is now large enough to be statistically significant,
the pattern of excess metallicity in host stars is fairly robust.

With biases for the most part ruled out, several competing theories have been devised
to explain the SWP metallicity enhancement trend. The main debate centers on whether
this pattern is primordial in origin. If so, one hypothesis maintains that planets
preferentially form around stars where there is ample heavy-element material. Hence,
those systems that contain above average metallicities will be more likely to spawn
planets. It is reasonable that the presence of metallic or rocky material in a
circumstellar disk might be favorable for planet formation;
\citet{1996Icar..124...62P} and \citet{1993prpl.conf.1061L} have outlined some of the
effects of disk composition on this process. Alternatively, some authors have
suggested that above-average metal abundances are not primordial, but rather, they
are due to the more dynamic phenomenon of planet or planetesimal accretion
\citep{2000ApJ...545.1064L,2001AJ....121..432G}. Accumulation of rocky material in
the outer layers of a star could produce an observable increase in surface
metallicity. In the absence of additional considerations, there is at present no a
priori reason to favor one hypothesis over the other. However, it is certain that an
adequate explanation of the planet/host-star metallicity trend will be crucial to
understanding the conditions under which planetary systems form and persist.

\subsection{The Possibility of Planetary Migration and Accretion}

The possibility that stellar abundances may be altered by the addition of
infalling rocky planet material is tantalizing from an observational
standpoint. Because this sort of enrichment is the result of a dynamical
process, as opposed to a preexisting condition, its signature should be
readily detectable in the surface abundances of stars, given the proper
diagnostics. Nevertheless, any feasible theory of metal accretion must
first answer the question of how such material could be delivered to the
stellar surface. Perhaps not coincidentally, some of the most metal
enriched systems exhibit strong evidence for {\em orbital migration}, in
which giant planets are forced inward toward their parent stars and settle
into short-period orbits. Yet, the idea of migration has taken
considerable effort to develop since it does not fit the ``established''
picture provided by our own solar system, where gas giants lie at
substantial distances from the sun, and large-scale inward orbital
migration has largely been ruled out \citep{1999AJ....117.3041H}.

Orbital migration due to circumstellar disk interaction is one explanation 
for the puzzling ``pile-up'' of planets at short distances from their host stars
\citep{1997iau.collq..163L,1997ApJ...491L..51L,1998ApJ...500..428T}. Most stars
harboring planets are assumed to be surrounded by protoplanetary disks until ages of
1-10~Myr \citep{1997ApJ...491L..51L}.  Before the time of dissipation, tidal forces
between a planet and the disk material in which it orbits may be responsible for
pushing the planet toward the central star. \citet{1997MNRAS.285..403G} linked this
idea to the observed metal enhancements in stars with close-in planets.  Citing
Lin, Bodenheimer \& Richardson's \citeyearpar{1996Natur.380..606L}
theory of planet-disk interaction, he hypothesized that high stellar metallicities
are directly related to planet migration, in that metal-rich dust ahead of the planet
is driven into the star. Other authors have similarly proposed that accretion 
of material is likely, on the grounds that the amount of time required for a planet
to migrate inward is close to the viscous evolution time of the disk
\citep{1997ApJ...491L..51L}. 

Even if pollution is not the main explanation for the high metallicities of SWPs,
more direct evidence for engulfment of planets has been reported in stellar
abundance anomalies. \citet{2001Natur.411..163I,2003A&A...405..753I} suggest that
the presence of $^{6}$Li in stellar atmospheres is an excellent tracer of rocky
material accretion, since this isotope should not usually be present in stars whose
convection zones once mixed material to depths with T$>2\times$10$^{6}$~K. They have
presented evidence for $^{6}$Li in the star HD~82943 (although the result is
disputed; see \citet[][]{2002MNRAS.335.1005R}), arguing that it must have been
delivered to the stellar surface at some point by infalling material. Further
support for pollution has been provided by differential Fe and Li abundance
enhancements in binary systems \citep[][and references therein]{2001A&A...377..123G,
2001ApJ...553..405L,2002ApJ...568..363Z}, although some of these results have since
been reconsidered \citep{2004astro.ph..3051D}. Binaries may provide an ideal
environment for identifying the signature of pollution, since the surface
composition of a pollution candidate may be easily compared to that of its
companion, which has presumably preserved the system's primordial abundances.
Nevertheless, these scenarios still cannot discriminate between accretion of
inward-migrating planets and other processes involving much smaller mixes of dust,
comets, or planetesimals as the source of metal-rich material.
 
\subsection{Theoretical Motivations} 

To make progress in determining whether abundance enhancements in stars with
planets and some binary system members is due to pollution by planetary material,
further observations and more constraining theories are required. We propose that
such constraints might be provided by detailed modeling of the interiors of stars
thought to have accreted high-Z material. We suggest that in depositing heavy metal
material in the interiors of stars, pollution should influence stellar structure
and evolution in a predictable way that is distinguishable from the behavior of
uniform-metallicity stars. Siess 
\& Livio (\citeyear{1999MNRAS.304..925S},\citeyear{1999MNRAS.308.1133S}) have
developed detailed models of the ingestion of massive planets by red giant and
asymptotic giant branch stars. Although some other groups have undertaken
hydrodynamical simulations of the stellar interior during consumption of a planet
\citep{2002ApJ...572.1012S,2001AAS...199.6111H}, the residual effects of small to
moderate amounts of metal enrichment on solar-like stars have not been considered
in detail. To understand the ramifications of pollution, there is a real need for a
precise description of the physical properties involved in planet (or otherwise
high-Z material) consumption.

The most quantitative analyses of pollution-induced metallicity enhancement have
focused on specific components of stars (e.g., the convection zone), rather than a
comprehensive treatment of the physics governing the stellar interior. Such
approaches are not self-consistent and may overlook potentially important phenomena
that arise in polluted stars. There is no guarantee, for instance, that a
metal-enriched convective zone will not alter the entire stellar temperature
structure or lead to runaway, nonequilibrium processes. The study of pollution's
effects on the stellar interior does have some history when it comes to our Sun. It
was suggested by \citet{1974ApJ...191..771J} that substantial contamination of the
Sun by infalling debris had masked the composition of the radiative solar interior,
following on earlier ideas by \citet{1939Obs....62..217H} and
\citet{1936MNRAS..96..559L}. Dirty solar models were developed by
\citet{1979A&A....73..121C} to study the effects on the solar neutrino problem.
These models treat the metal-enriched convection zone (both its lower and upper
boundaries) in a very simple manner. Recent dirty solar models by
\citet{2002ApJ...576.1075W} use the same approach, which brings the star to a new
thermal equilibrium by adjusting a free parameter, like the mixing length, that is
not constrained by observations. When applied to stars other than the sun, the
simple approach of enhancing the metallicity of the surface layers does not allow
the star to adjust to a new thermal equilibrium, as Ford, Rasio \& Sills
\citeyearpar{1999ApJ...514..411F} pointed out. \citet{2003ApJ...596..496D} have
recently presented similar polluted stellar evolution models that do not describe
the physical changes in the interior structure.

Fully self-consistent polluted stellar evolution models for solar-type stars have
not been discussed extensively in the literature. To this end, we have modified
an existing accurate stellar evolution code to allow for discontinuities in
metallicity inside stars. We aim to examine the effects of differential metal
enrichment on stellar structure and evolution, when applied at times close to and
after zero-age main sequence (ZAMS). By incorporating a pollution scheme directly
into the evolution program, we allow stars that accrete heavy material
to undergo a thermal relaxation process prescribed solely by the physics already
built into the code, thereby producing one continuous evolutionary model. Our 
motivation for carrying out these simulations is in the
possibility that many stars might have consumed planets during different stages
of their evolution; mostly before settling on the ZAMS, but not restricted to
that. Since current evidence is limited to stars similar to the sun, we
will consider only stellar models in the 0.8$-$1.4~M$_{\odot}$ range. Such stars
have surface convection zones and mixing layers, which play a crucial role in the
changes we study. In subsequent sections, we discuss the main features of the
resulting polluted stellar evolution models and provide a comprehensive set of 
evolutionary tracks.
 
\section{Adding ``Metals" to a Surface Convection Zone}

Prior to detailed numerical modeling, we highlight the dominant physical changes
expected subsequent to metal enhancement events in convection zones of stars. As
discussed in \S 3.2, the planetary material is assumed to be confined within the
outer convective region of the accreting star.  This leads to a discontinuity
in relative metal content (by mass), Z, and hence molecular weight, at the
convection zone boundary.  Assuming Kramer's type bound-free opacity sources
typical of lower main sequence stars with Z$\gtrsim$10$^{-4}$, we have
$\kappa\propto$Z(1+X)$\rho$T$^{-3.5}$, where $\kappa$ is the opacity, and X is the
relative hydrogen content by mass. After enrichment in a convection zone, in which
the metal fraction is raised from some value Z to Z+$\Delta$Z, X will decrease to
X($\frac{1-Z-\Delta Z}{1-Z}$). As long as Z is not unreasonably large, $\kappa$
will then increase, based on its Z(1+X) dependence. Thus, there should also be a
change in opacity at the convection zone boundary. Updated data from the OPAL
opacities \citep{1996ApJ...464..943I} suggests that this change should be {\em
particularly} pronounced near the base of the stellar convection zone, since the
opacity is quite sensitive to the Fe abundance at the $\sim10^{6}$~K characteristic
temperatures here. The requirements for stability at this location will presumably
be affected by the discontinuities in molecular weight and opacity, since the
radiative temperature gradient, $\nabla_{\rm rad}$ is directly proportional to
$\kappa.$ As established by the Schwarzschild criterion, the onset of convection
occurs where the radiative temperature gradient exceeds the adiabatic temperature
gradient, $\nabla_{\rm ad}$. The increase in $\nabla_{\rm rad}$ at the bottom of
the convection region should cause convective instability to extend to lower depths
within the star.  The related phenomenon of a more shallow convection zone in the
absence of high metal content is indeed observed in stars with overall low
metallicity \citep{1982A&A...115..357S}.

Pollution-induced changes should not only be isolated to the lower portion of the
convection zone. Qualitatively, elevated opacities will restrict the amount of
radiation passing from the inner to the outer regions of the star. In general, stars
of overall high metallicity occupy main sequence positions on the H-R diagram that
are lower in both luminosity and temperature than their low-metallicity
counterparts. We expect roughly similar behavior for polluted stars, as they undergo
initial adjustments after addition of high-Z material. But since the mass of
accreted metals and size of the convection zones are small compared to the overall
dimensions of the stars under consideration, pollution may not have any long term
effects on the central nuclear burning, and thus the luminosity.

The persistence of elevated opacity in the convection zone should, however, indefinitely
maintain a lower effective temperature (due to lower flux) than that of the
corresponding unpolluted star. For negligible changes in luminosity, this effect causes
an increase in the stellar radius. The expansion can be understood as an adjustment
necessary to restore radiative equilibrium in the outer regions of a polluted star by
increasing the area over which energy may escape outward.

In summary, the general physical outcomes to be expected from metal pollution
include both shifts in lower convection zone limit and changes at the upper
boundary, due to enhanced opacity.  Such enriched convection zones are
hydrostatically stable, and the adjustment of stellar structure is a gradual
quasi-equilibrium process (D.~Sasselov \& B.~Hansen 2003, private communication).

\section{Polluted Stellar Models: The Modified Computation}
\subsection{The Baseline Stellar Evolution Code}

The baseline code is the same as that used in our recent work \citep[][hereafter
CS02]{2002ApJ...569..451C}, and based on the Princeton Stellar Evolution Code
(Sienkiewicz, Paczynski \& Ratcliff \citeyear{1988ApJ...326..392S}). The program solves
the equations of stellar structure on a one-dimensional stellar mass grid, making use of
the OPAL radiative opacities \citep{1996ApJ...464..943I, 1995aapn.conf...31R}, Livermore
Laboratory equation of state data (Rogers, Swenson \& Iglesias
\citeyear{1996ApJ...456..902R}), and nuclear reaction rates from Bahcall, Pinsonneault,
\& Wasserburg \citeyearpar{1995RvMP...67..781B}. Molecular and grain opacities supplied
by \citet{1994ApJ...437..879A} are also employed in lower temperature regions of the
opacity tables, and equations of state correspond to an ideal or partly ionized gas. In
addition to tracking temperature, luminosity, and density evolution, the code traces
abundances of H, He3, He4, N14, O16 and O17 as produced by hydrogen burning in the pp
chain and CNO cycle; in this work, we consider all elements heavier than helium as part
of the stellar ``metal'' content. Diffusion of helium and other elements has not been
incorporated into the calculations. Other conditions modeled in the code include slow
rigid-body rotation, as well as a simple overshooting scheme. The number of mass points
in the grid is adjustable, and we find that it must be maintained at a minimum of 4000
to provide satisfactory resolution.  Limitations in the equation of state tables also
restrict models to $\gtrsim$0.87~M$_{\odot}$. We are able to verify the general accuracy
of the code in the ``allowed'' parameter space by comparing an evolutionary track at
solar mass and metallicity with the sun's known temperature and luminosity. Good
agreement is obtained, and we attribute the slight difference ($\sim0.2\%$) between
theoretical and observed parameters to neglect of helium diffusion. We have also run
comparisons with the Yale code (see CS02).

\subsection{Assumptions of Star/Planet Interaction}

 To trace the effects of heavy-element accretion due to consumption of 
planets, we have modified the baseline stellar evolution code to allow 
for non-uniform metallicity distributions within stars. Our models
proceed on several assumptions. First, we require that material from
inward-migrating planets enters the stellar envelope and dissolves
completely within the convection zone, contributing negligible
contamination to the inner regions of the star.  While there remain many
questions about the dynamics of orbital decay both before and after
entrance into a host star, recent calculations indicate that complete
disintegration of a planet within the convection zone is realistic for a
range of planet masses, compositions and orbital time scales.
\citet{1998ApJ...506L..65S,2002ApJ...572.1012S} have evaluated the
feasibility of this scenario with hydrodynamical simulations of giant
planet accretion.  Testing planet consumption in stars in the
neighborhood of 1~M$_{\odot}$, they find that pollution is capable of  
producing observable surface metallicity enhancements, but the percentage
of planetary mass dissolved within the convection zone is highly
dependent on the structure of both star and planet.  In addition to gas  
giants, smaller rocky planets and planetesimals have also been cited as
potential sources of pollutant material \citep[][and references
therein]{2001A&A...377..123G}. Because of the current lack of constraint
on amount and type of debris that could be accreted onto stars, pollution
is best characterized as a change in metallicity; our models make no
prior assumptions about the nature of the polluting planets, except to 
specify the mass of heavy element enrichment.  Furthermore, we restrict
the physics to scenarios in which the accreted metal content is mixed 
evenly throughout the convective region. We note that the extent of  
observable surface metallicity enhancement is governed by the structure
(i.e., depth, density) of the stellar convection zone. 

 An additional consideration relevant to the evolution of polluted stars
is the {\em time} of planet accretion.  Since the mechanisms of orbital
migration are not well understood, no particular time appears most
favorable for planetary migration and consumption. A rough lower bound is
offered by evidence that any {\em observable} pollution takes place in stars that have
already reached the main sequence, because the large convection zones of
pre-main-sequence stars would lead to considerable dilution of the polluting
material \citep{2000ApJ...545.1064L,2001AJ....121..432G}.  Consequently, the models 
produced for this analysis disregard stars that reside on the Hayashi
pre-main sequence tracks. In substantially evolved stars, on the other
hand, there is no reason to discount the possibility of metal enrichment.
Hence, the time scale pertinent to pollution runs from the zero-age main
sequence to the subgiant stage.  By specifying the stellar age at time
of enrichment along with the amount of metallicity enhancement, we are
set to describe stellar pollution as a simple two-parameter process.

\subsection{Implementation of the Pollution Scenario}

 Under the above considerations, our adaptation of the Princeton Stellar Evolution
Code according to pollution theory leaves most of the original physics unaltered. All
models employ mixing length theory to compute the convective envelope, incorporating
the solar mixing length value of 1.69 \citep{1997ApJ...484..937G}. In addition,
helium content (Y) is varied in accordance with a suitable linear enrichment law,
with $\delta$Y/$\delta$Z=2.5, and we have chosen as a zero point the well-known solar values,
(Y$_{\odot}$, Z$_{\odot}$)=(0.2741, 0.0200) \citep{1997ApJ...484..937G}. Hydrogen, X,
is dependent on metallicity and given by the usual equation $X+Y+Z=1$. Primary
modifications involved changing the program to permit variation of metallicity with
stellar radius. We parameterize pollution by $\Delta$Z, the change in convection zone
metal content due to planet consumption, and $t$, time of enrichment (Gyr). Pollution
is carried out by uniformly increasing the value of the relative metal abundance (Z)
within the confines of the convection zone at the specified time. The code determines
these boundaries and the corresponding convection zone mass, M$_{\rm cz}$, through
the Schwarzschild stability criterion (i.e., radiative temperature gradient equals
adiabatic temperature gradient).  Although uniform enhancement of convection zone
metallicity may not be accurate insofar as mixing requires substantial time, the
relevant time scales for planet consumption, as well as radial mixing and meridional
circulation are short enough compared to the overall evolutionary time steps that
they may be neglected. The former is approximately $\lesssim$10$^7$~sec (as found
with our code), and the latter is up to $\sim10^3$~yr for the most slowly rotation
stars \citep[see][for a discussion of meridional circulation in the
sun]{2002ApJ...570..865B}.

\begin{figure}[!h]
\begin{center}
\plotone{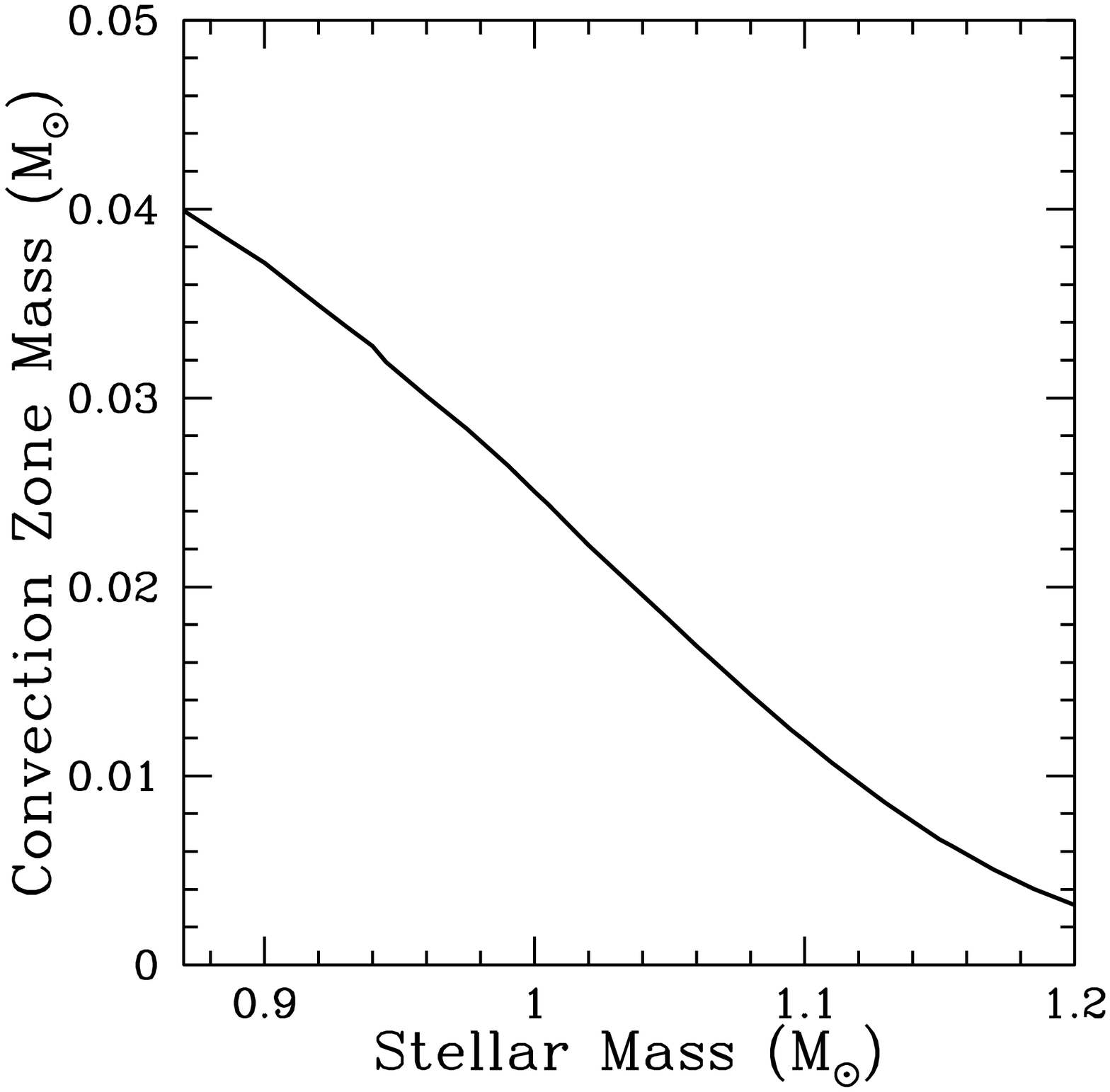}
\end{center}
\figcaption{Theoretical predictions for stellar convection zone mass as a
function of total stellar mass for stars at age 10$^{5}$ years.\label{conmass}}
\end{figure}

 During and after planet accretion, several conditions influence $\Delta$Z, the value
of metal enhancement within the convection zone, as well as the subsequent overall
metallicity distribution. Convection zone depth is one crucial factor, and its
thickness as a function of total stellar mass is well known. In stars above
1.3~M$_{\odot}$, this region is nearly non-existent, while for stars below
1~M$_{\odot}$, it deepens rapidly. Yet for masses in the range 1.2--1.5~M$_{\odot}$,
where the convection zone is predicted to be very thin, further mixing is observed,
as in the ``Lithium dip'' present in Hyades stars \citep{1986ApJ...302L..49B}.  This
phenomenon has largely been attributed to stellar rotation
\citep{2002ApJ...565..587B}. But since it is not well accounted for in standard
stellar evolution theory, we only employ the code in modeling pollution for stars
below 1.2~M$_{\odot}$. In this range, convection zone depth monotonically decreases
with increasing mass. As Pinsonneault, DePoy
\& Coffee \citeyearpar{2001ApJ...556L..59P} and others point out, pollution
should be most easily observable in stars at the high-mass end, because there will be
little dilution of metal enrichment within their thin convection zones. Since we have
chosen to give $\Delta$Z a priori for each model, without regard to size of the
convection zone, this simply means that the corresponding amount of pollutant
material needed will be smaller in models of higher overall mass. For a given
$\Delta$Z, conversion from metal enrichment value to pollutant mass is carried out by
taking into account the mass enclosed within the convection zone limits, M$_{\rm
cz}$. Using the code, we derive this quantity as a function of overall stellar mass,
as shown at age log(t)=5 (due to the code time step, we cannot produce the zero-age
convection zone) in Fig.\ \ref{conmass}. Ignoring additional mixing evident in the
``lithium dip," this is consistent with the results of \citet{2001ApJ...555..801M}.

Of lesser importance in determining the metallicity distribution are progressive
shifts in the convective boundary. Although the initial amount of metal throughout the
entire star (Z) and within the convection zone (Z+$\Delta$Z) is fixed in each
model, expansion of this region leads to slight dilution of the enriched material.
Slow changes in the boundary are typical during stellar evolution but, as
discussed in \S 2.2, they should be more pronounced after addition of
high-metallicity material, as a result of increased opacity. The modified version
of the code has been designed to recalculate the metallicity-dependent opacities
and equations of state at each location in the stellar grid where Z differs
significantly from that of neighboring points. Interpolation is performed between
(X,Z) entries in the equation of state and opacity tables to provide the most
accurate values.  When pollution induces a downward shift of the convection zone
boundary, the metal-enhanced region is mixed with the layer of elements below.  
As during the initial pollution event, re-mixing of the convection zone contents
takes place during each single time-step of the simulation. This deepening of the
convection zone is followed by the code and tends to reduce the observed
metallicity enrichment due to pollution.

\begin{figure}[!h]
\begin{center}
\plotone{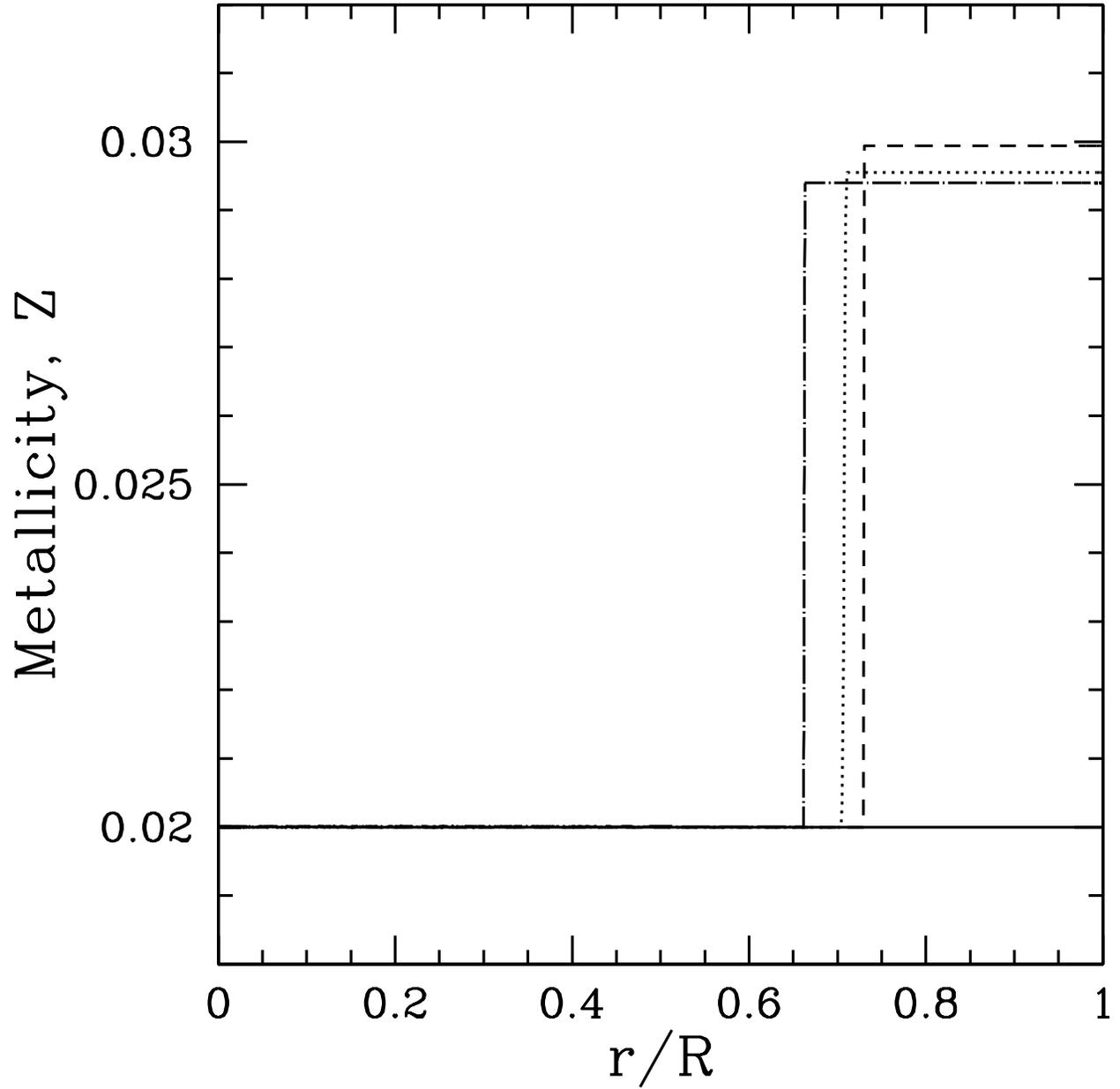}
\end{center}
\figcaption[Radial Metallicity Profile]{For a 1.0~M$_{\odot}$ star polluted from
Z$_{\circ}$=0.02 to 0.03, Z is shown as a function of radius at ages
t=10$^{5}$ (solid line), 10$^{5.2}$ (long dashes), 10$^{9}$ (dots), and 10$^{10}$~Gyr
(dot-dash)\label{dilution}}
\end{figure}

 Overshooting is an additional effect that counteracts the relative metal enrichment
generated by pollution in the upper layers of the star. Within the evolution code, the
top and bottom of the convection zone are determined with the Schwarzschild criterion;
in reality, we expect substantial quantities of higher metallicity material to be
transported beyond the convective boundaries. A simple overshooting model was
implemented in the baseline code to account for this phenomenon. During the majority of
our pollution simulations, enriched material is not only circulated within the
convection zone, but is also efficiently mixed throughout the atmosphere above.
Combined with the consequences of a deepening of the convection zone, or an initially
large convective depth, overshooting has the potential to reduce $\Delta$~Z from its
initial value. In Fig.\ \ref{dilution}, we show radial metallicity profiles for a
1~M$_{\odot}$ at several different times, both before and after pollution; gradual
dilution due to these effects is evident.  Ignoring any mass loss or mixing in
semi-convective regions, no other processes contribute to significant dilution of the
higher metallicity region. We confirm the accuracy of the modified code in the limiting
case $\Delta$Z=0 by showing that it produces the same evolution tracks as the original
program.  The results of {\em polluted} evolution, however, are much more difficult to
verify, since there are no precedents for this sort of stellar model in the literature.

\section{Pollution: General Trends}
\subsection{Convection Zone Changes}

\begin{figure}[!p] 
\begin{center} 
\plotone{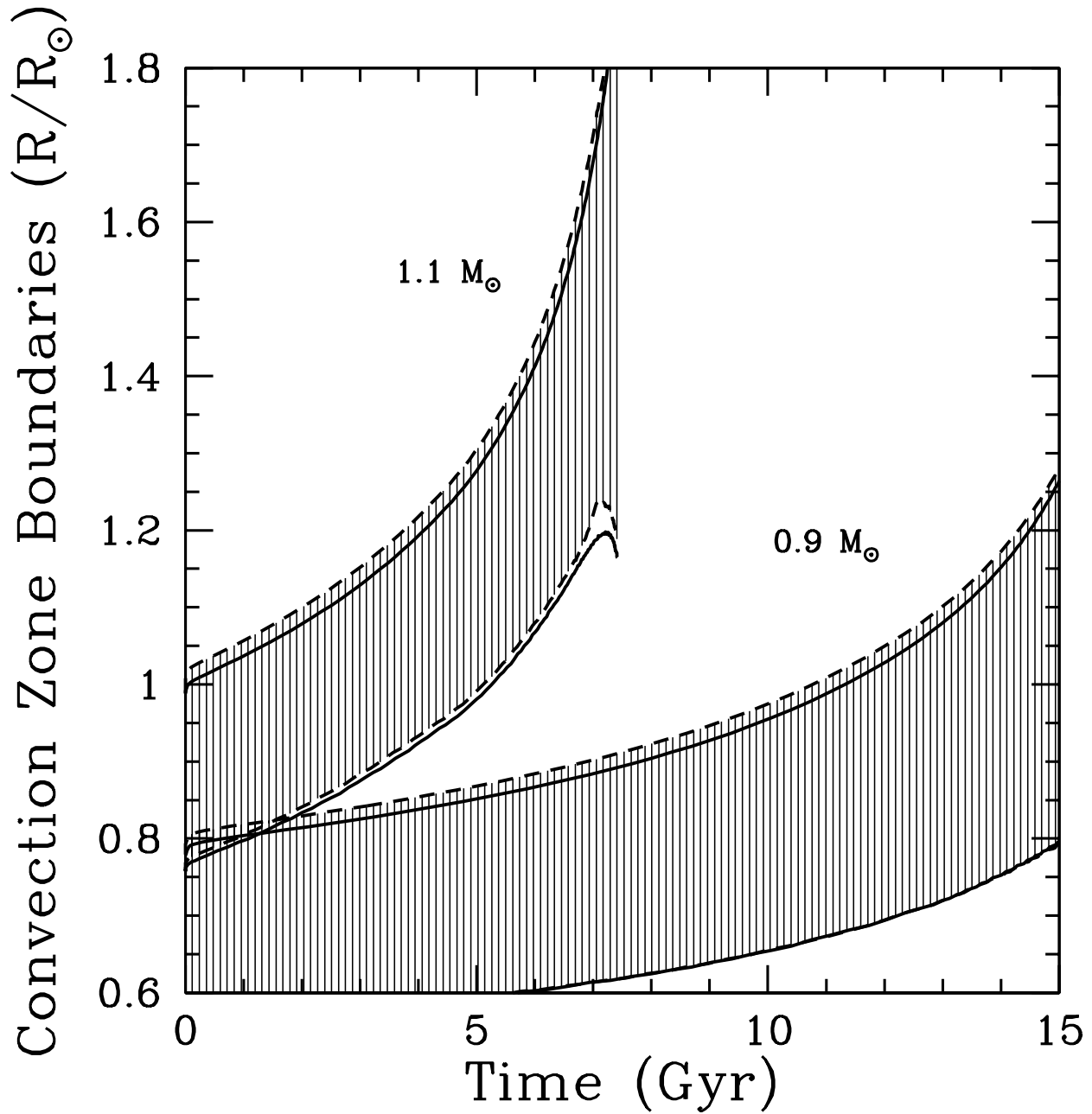} 
\end{center} 
\figcaption[Radial boundaries of the polluted and
unpolluted convection zones]{Radial boundaries of the stellar convection zone versus
time are plotted for polluted and unpolluted models with X=0.7, Z=0.02, and masses
1.1~M$_{\odot}$ and 0.9~M$_{\odot}$. Convection zone boundaries of polluted models
are marked with dashes and were produced with $\Delta$Z=0.01; boundaries for the
unpolluted models are marked as solid lines.  Vertical shading shows the extent of
each convection zone for clarity; the outer radius of the star is not drawn, because 
it is within 0.5\% of the convective region upper boundary. 
\label{convrad}} 
\end{figure}

 Before analyzing full stellar models, we investigated differences in the
convection zone boundaries of stars with and without pollution.  As previously
discussed, we expect a deepening of the lower convection zone boundary, due to
increased opacity.  We have modeled both the mass contained within polluted and
unpolluted convection zones, as well as the radii of their boundaries.  As seen
in Fig.\ \ref{convrad}, the top of the convection zone extends outward nearly to
the stellar surface. Upon pollution with planetary material, a star responds by
increasing its radius, whereas the inner boundary of the convection zone remains
at virtually the same absolute distance from the core.  The combination of these
two effects leads to a relative deepening of the convection zone.

\begin{figure}[!p]
\begin{center}
\plotone{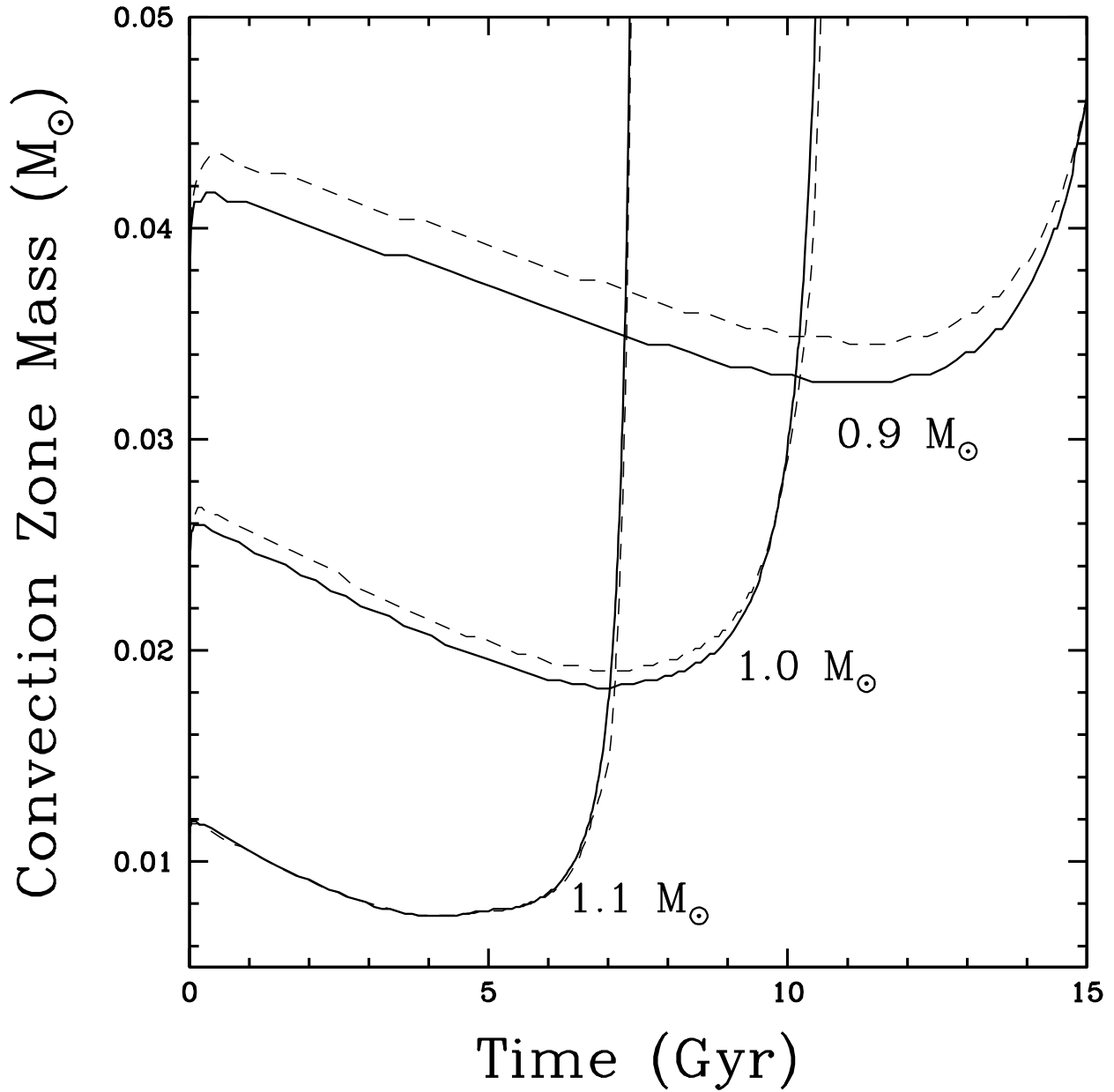}
\end{center}
\figcaption[Masses of polluted and unpolluted convection zones]{Mass of
the stellar convection zone versus time is plotted for polluted
and unpolluted models with Z=0.7, Z=0.02, and masses  1.1~M$_{\odot}$,
1.0~M$_{\odot}$, 0.9~M$_{\odot}$. Polluted models are marked with dashes and were
produced with the specification $\Delta$Z=0.01. Small ridges on the curves are
due to limited mass resolution and are not true features of the convection zone
evolution. \label{convmass}}
\end{figure}

 The behavior of the convection zone {\em mass} is also in line with our     
theoretical predictions.  Depicted in Fig.\ \ref{convmass}, it displays a marked increase
in stars of significant convective depth (i.e., those with 1.0~M$_{\odot}$,
0.9~M$_{\odot}$, and lower).  The enclosed amount of mass is linked with the
radial changes, but tied to the stellar density profile as well. Thus, in the
1.1~M$_{\odot}$ model, the polluted convection zone is actually a bit smaller  
than its unpolluted counterpart; this is the outcome of the larger
stellar radius and correspondingly decreased density near the surface.  In
addition, there is an obvious expansion of the convection zone as model stars
approach the subgiant branch; this can be seen if both Fig.\ \ref{convmass} and
Fig.\ \ref{convrad}.

\subsection{Polluted Stellar Evolution Tracks}

 We have calculated polluted stellar evolution tracks for a variety of
stars from M=0.9--1.2~M$_{\odot}$ and Z=0.01--0.03.  For models below
0.85~M$_{\odot}$, the convection zone becomes quite deep, and
effects of pollution are not expected to be significant. For purposes of   
comparison, most tracks were calculated assuming pollution at the zero-age 
main sequence. However, an additional set of models was produced to explore
the outcome of pollution at much later times, from $2\times10^{5}$~yr to
9~Gyr. We now discuss several notable trends present among the polluted
stellar evolution tracks.

\begin{figure}[!ht]
\begin{center}
\plotone{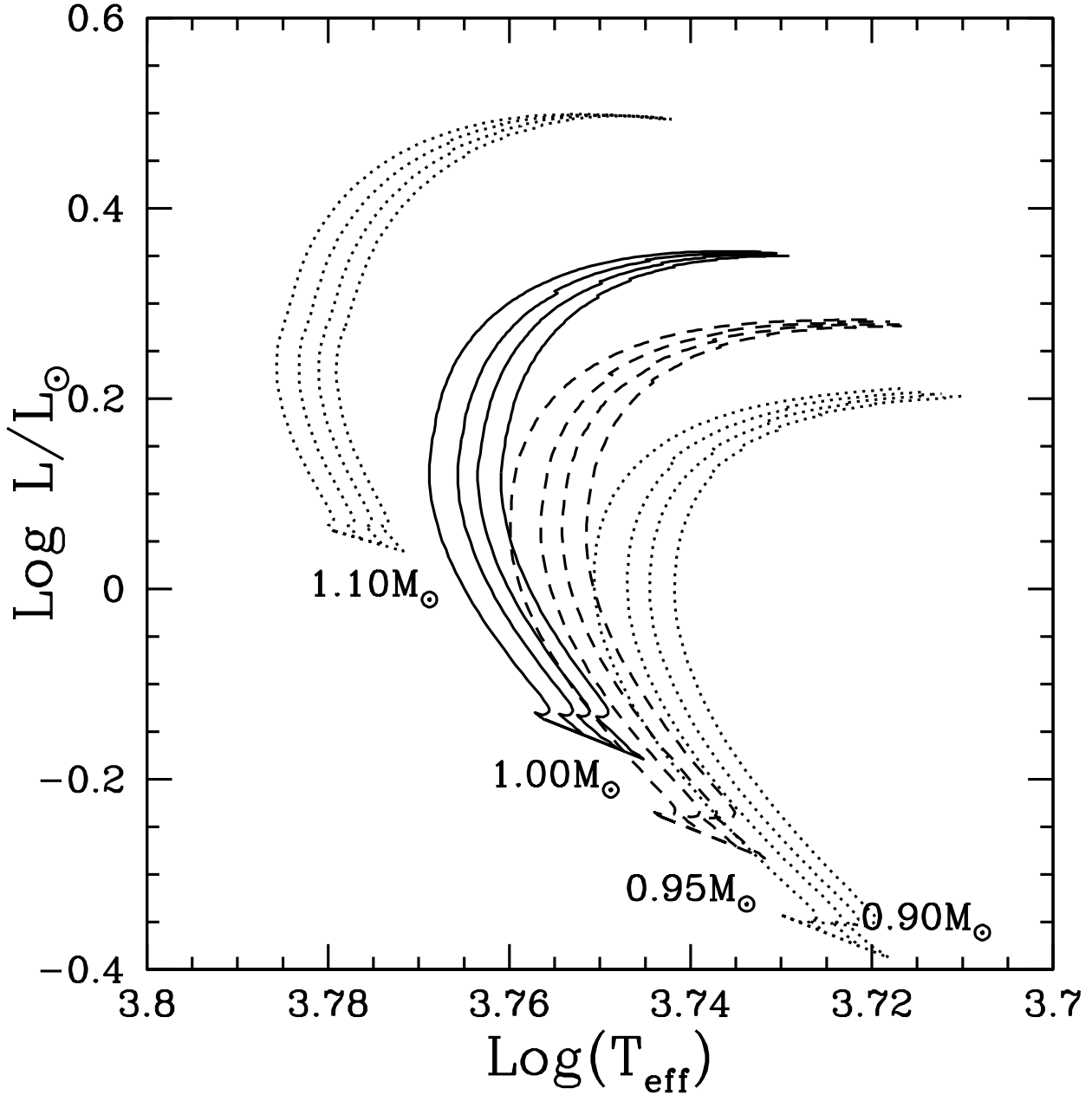}
\end{center}
\figcaption[Polluted stellar evolution tracks for Z=0.02 and varying mass]{Polluted
stellar evolution tracks for a fixed metal fraction of Z=0.02, and various masses.
Larger amounts of pollution ($\Delta$Z) shift the tracks
toward lower temperature.  The values of $\Delta$Z included on the plot are
$\Delta$Z=0.0, 0.005, 0.01, and 0.015 for all masses. It is evident that the
temperature shift due to pollution is roughly proportional to $\Delta$Z. As the
mass increases and the convection zone shrinks, this shift decreases, becoming
nearly negligible for a 2.0~M$_{\odot}$ model. Ages range from 6.2--15.8~Gyr. 
\label{zconst}}
\end{figure}

\begin{figure}[!ht]
\begin{center}
\plotone{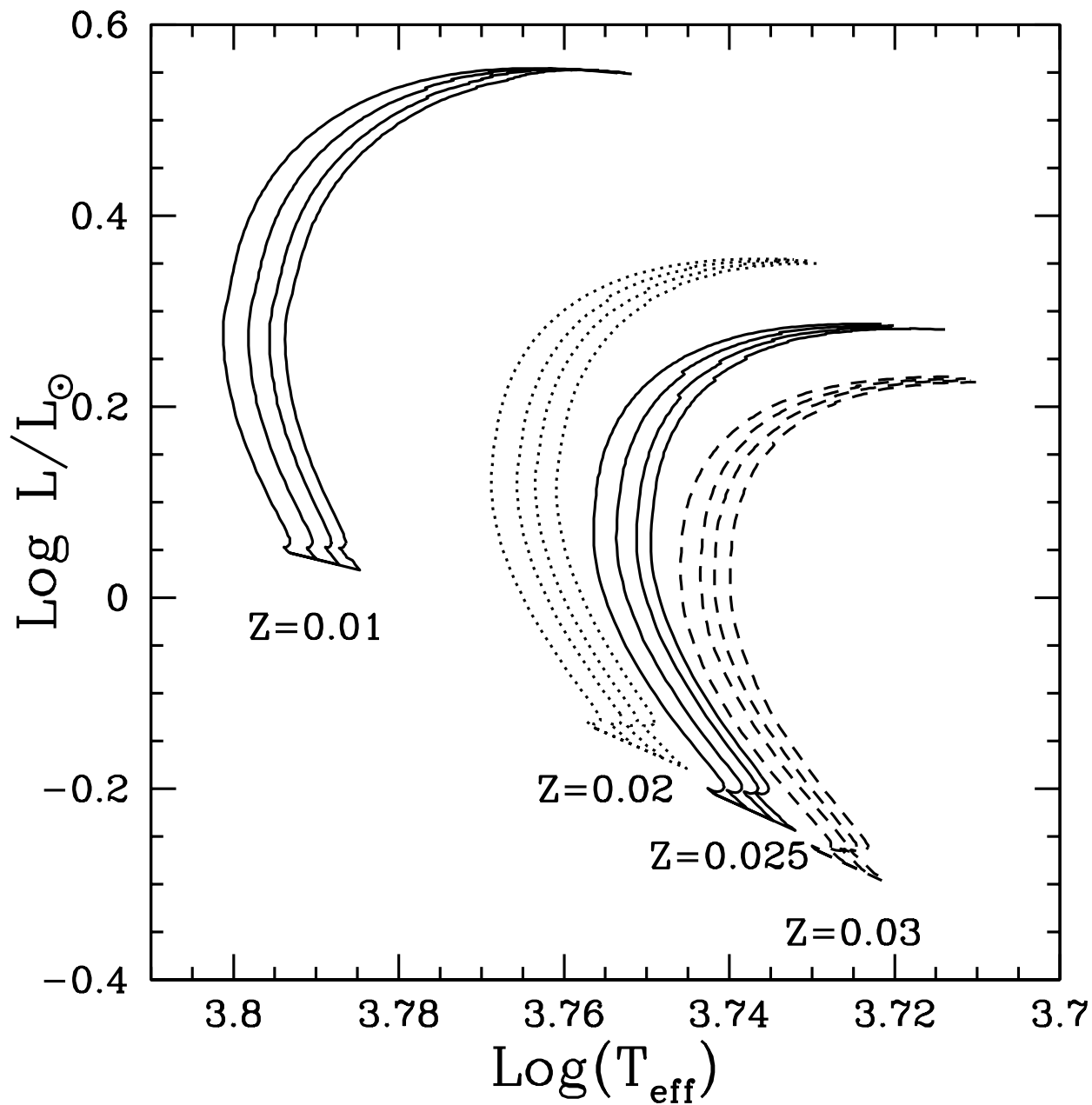}
\end{center}
\figcaption[Polluted stellar evolution tracks for M=1~M$_{\odot}$and varying
metallicity]{Polluted stellar evolution tracks for a 1~M$_{\odot}$ star, and several
different metallicities. $\Delta$Z values included for each metallicity on the
plot are $\Delta$Z=0.0, 0.005 ,0.01, and 0.015.  In all cases, temperature
decreases with increasing $\Delta$Z. The locations of sets tracks are affected
by metallicity changes, but the magnitudes of smaller shifts solely due to
pollution show little variation. As with the fixed-metallicity models, the
temperature shifts scale with amount of pollution. Tracks cover a range of
approximately 7.2--14.4~Gyr. \label{mconst}}
\end{figure}

 As seen if Figs.\ \ref{zconst} and Fig.\ \ref{mconst}, pollution causes evolution tracks to
shift downward in temperature.  This shift is the main feature of the thermal adjustment
process resulting from the addition of heavy-metal material to the outer zones of a star.  
The duration of adjustment is given roughly by the thermal relaxation timescale
\citep[see][for a detailed discussion of this phenomenon as expected in accretion of material
onto the secondary star of a cataclysmic variable system]{1999MNRAS.309..245S}, so the full
extent of the temperature shift is completed over only one timestep within the evolution
models. Our results are consistent with the models of Dotter \& Chaboyer, which 
require that additional metals be introduced over {\em several} timesteps; they find that 
polluted stars are hotter than unpolluted stars with the same mass, age, and metallicity.

For all test cases from M=0.9--1.2~M$_{\odot}$ and Z=0.01--0.03, the temperature
shift is roughly proportional to the parameter $\Delta$Z. While the
proportionality factor varies with mass, this can be attributed to the fact that
size of the convection zone is directly related to stellar mass. For constant
values of $\Delta$Z, the physical amount of metals added to the convection zone
will depend on the mass of the convection zone. Thus, for higher masses, the
convection zone is smaller, less pollutant material is introduced, and hence the
effect is smaller. Aside from this trend, the only factor determining the
magnitude of temperature shifts from model to model seems to be the value of
$\Delta$Z. The remarkable consistency in this result could be useful in
discriminating between pollution or lack thereof, if the polluted version of the
code is applied to stars known to harbor planets.

\begin{figure}[!p]
\begin{center}
\plotone{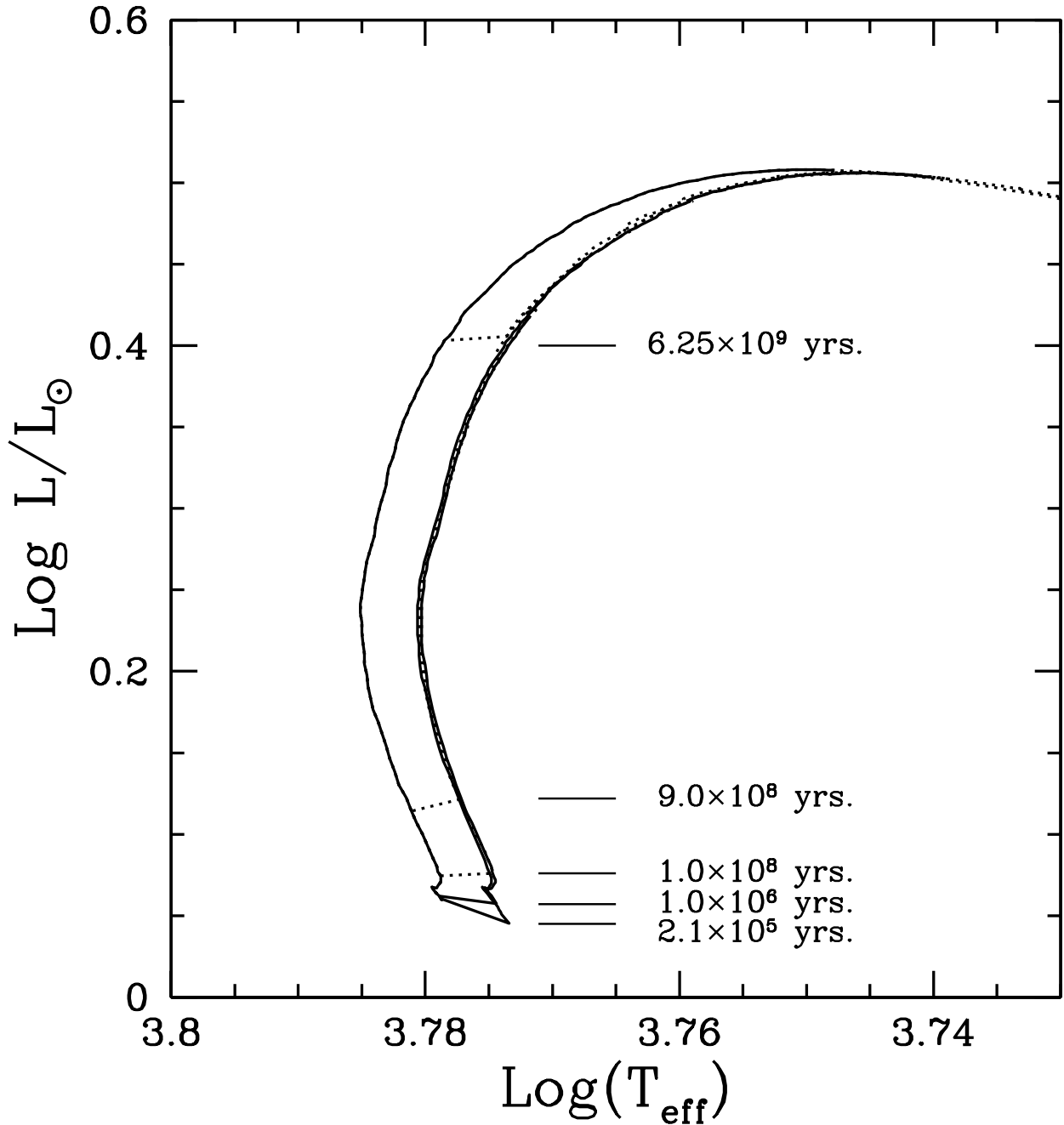}
\end{center}
\figcaption[Pollution at varying times ]{Results of varying the {\em time} of
pollution in a 1.1~M$_{\odot}$ star with Z=0.02 and $\Delta$Z=0.01. We have
calculated models for a variety of pollution times, and the resulting tracks are
coincident, regardless of the age when the metallicity is increased (assuming that
it is after the ZAMS time and before the subgiant stage). The plotted models follow the
unpolluted track until pollution occurs, at which point they cross over to the polluted
track (lower temperature). Age at crossover is noted on the plot. \label{age}}
\end{figure}

 Another feature verified by the computer modeling is that pollution-induced
shifts in evolutionary tracks are independent of when pollution happens, for times
after the late as the ZAMS age and before the subgiant stage. Polluted models were
calculated for a 1.1~M$_{\odot}$ star starting at
solar metallicity, and polluted by $\Delta$Z=0.01. When pollution was delayed
until the arbitrary times of $2.1\times10^5$, $1.0\times10^6$,
$1.0\times10^8$, $9.0\times10^8$, and $6.25\times10^9$~yr, the polluted and
unpolluted portions of each track remained coincident with each other (see
Fig.\ \ref{age}).  We conclude that a polluted star will indeed reach the same
equilibrium state, regardless of whether high-metallicity material is
acquired immediately, or at a later time when the star has evolved a bit.
This result indicates that pollution of the surface convection zone has ltitle to no 
effect on the central burning process, and reinforces the simplicity of modeling 
polluted stars, since we are now able to omit the time of pollution as a free parameter. 

\subsection{Application to HD~209458}

 As a test of the possible applications to our modified stellar evolution code, we
consider the case of the star HD~209458, host to a transiting planet (see CS02).  At
[Fe/H]=0.0, the star does not appear to be particularly enriched with metals. Nevertheless,
if we assume that the majority of stars with planets fall on an enhanced metallicity
distribution a full 0.2 dex higher than ``ordinary'' solar neighborhood stars, then it
is not unreasonable to suppose that HD~209458 was polluted from an original value of
[Fe/H]=-0.2. There is further uncertainty, however, in the conversion from initial
[Fe/H] to initial Z, since the {\em iron} content of pollutant is not known.  
We explore the range of possibilities by considering two cases: 100\%
iron pollution, and a mix of metals in the same proportions as found in the sun (in which iron
comprises 7.18\% of metals by mass; \citet{1993QB450.O76......}). These correspond to initial Z
values of 0.019 and 0.013, respectively. The amounts of metal pollution $\Delta$Z needed to 
raise the initial Z to 0.02 on the surface are then 0.001 and 0.008 (allowing for some dilution 
of the metals due to overshooting and expansion of the convection zone).

\begin{figure}[!h]
\begin{center}
\plotone{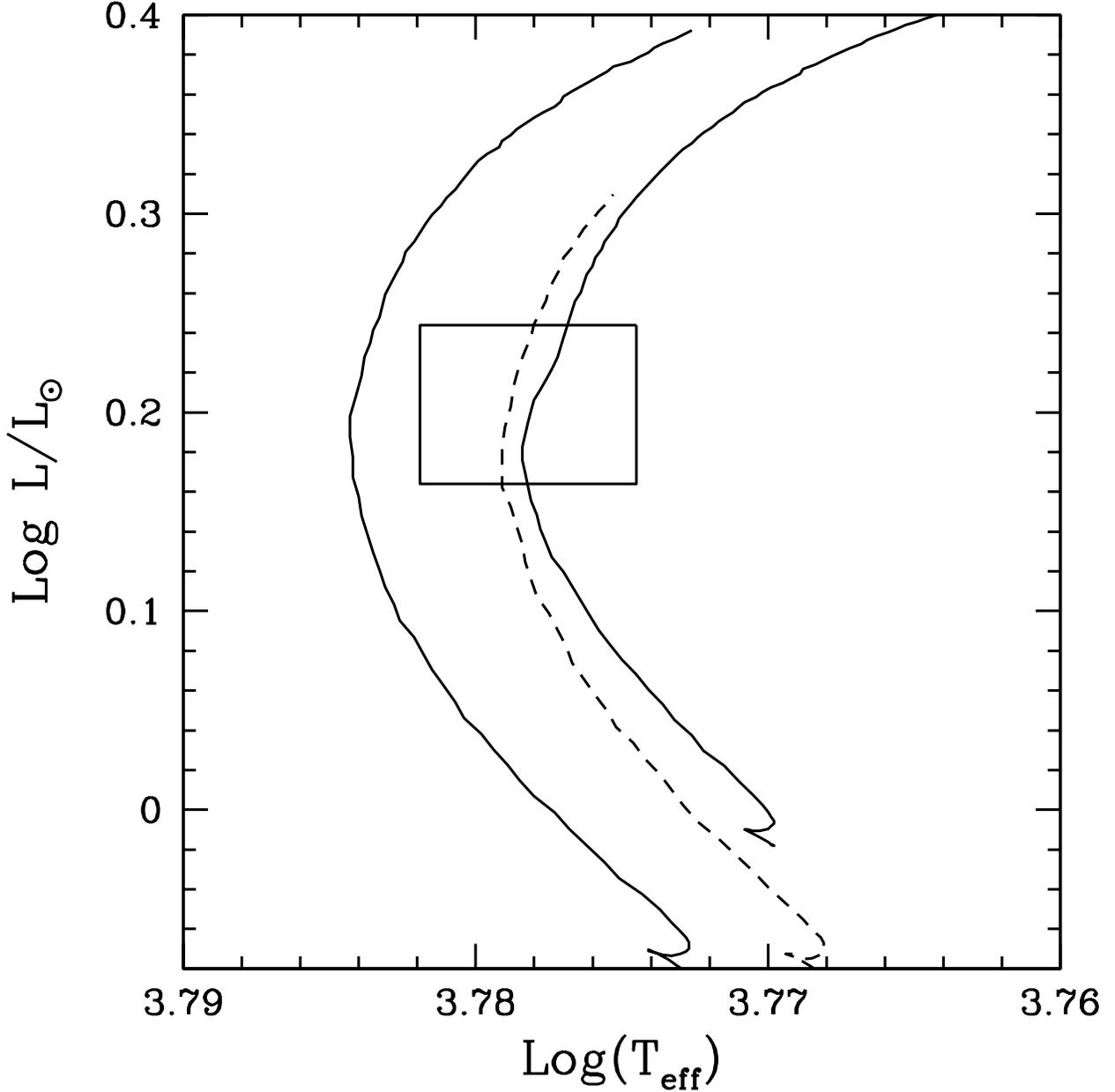}
\end{center}
\figcaption[Polluted stellar evolution tracks for HD~209458]{We have computed polluted
and unpolluted evolutionary tracks for the star HD~209458, whose
temperature-luminosity error box is shown in the center of the plot. 
A polluted stellar evolution track with parameters M=1.06~M$_{\odot}$, Z=0.019, $\Delta$Z=0.001, 
and 100\% iron pollution is shown with a dot-dash pattern.
The dashed curve illustrates the best-fit polluted track for HD~209458 when the accreted material 
is assumed to contain only metals in solar proportions; it is achieved with the values
M=0.97~M$_{\odot}$, Z=0.013, and $\Delta$Z=0.008. The solid track within the
box shows the best unpolluted evolutionary model for Z=0.02, at M=1.06~M$_{\odot}$.  For
comparison, we have also plotted an unpolluted track of M=0.97~M$_{\odot}$, Z=0.013,
which appears to the left of the error box. \label{hd209}}
\end{figure}

For the unpolluted case (Z=0.02, $\Delta$Z=0.0) as well as the two polluted cases, we 
have computed evolution tracks to match HD~209458's observed temperature of 
log(T$_\mathrm {eff}$)=3.778 and luminosity log(L/L$_{\odot})=0.208$.
The best-fit models are illustrated in Fig.\ \ref{hd209}. Further
models were fit to the edges of the T$_\mathrm {eff}$-L error box to determine the
range of possible masses and ages for the two scenarios, as given in Table 1. When 
metallicity is also varied according to [Fe/H]=$0.0\pm0.2$ (or Z=$0.02\pm0.005$), the 
mass uncertainty increases to approximately $\pm0.10$.

\vspace{0.3cm}
\begin{table}[!h]
\begin{center}
\begin{tabular}{ccccc}
\hline
\hline
 & \phm{rr}Mass & \phm{rr}Age  & \phm{rr}Initial\phm{r} & Initial Convective \\
Model & \phm{rr}(M/M$_{\odot}$)& \phm{rr}(Gyr) & \phm{rr}[Fe/H]\phm{r}& Mass 
(M/M$_{\odot}$)\\
\hline
\\
Unpolluted&  $1.06^{+0.03}_{-0.02}$& $5.7^{+1.5}_{-2.5}$&\phm{r}$0.0$& $1.8\times10^{-2}$ \\
\\
Polluted: Metals in & $0.97\pm0.02$& $6.6^{+1.5}_{-2.0}$& $-0.2$& $1.5\times10^{-2}$ \\
Solar Proportions & & & & \\
\\
Polluted: 100\% Fe & $1.06^{+0.03}_{-0.02}$& $5.6^{+1.5}_{-2.5}$& $-0.2$& $1.8\times10^{-2}$ \\
\\
\hline
\hline

\end{tabular}
\caption[Parameters for HD 209458]{Best-fit parameters for HD~209458 based on 
log(T$_{\rm eff})$=$3.778\pm0.004$,\\ log(L/L$_{\odot})=0.208\pm0.040$, and 
Z$_{\rm surface}=0.02\pm0.005$ (see CS02).}
\end{center}
\end{table}

 Assuming the solar mixing length, 1.69 \citep{1997ApJ...484..937G} and helium
(X=0.70), we find a reduction in mass by 0.09~M$_{\odot}$ and increase in age by
1.4~Gyr from the best-fit unpolluted values. For the polluted model with solar metal
composition, we find M=0.97~M$_{\odot}$ and age 6.6~Gyr. This value is at the very edge
of allowable masses found by CS02, taking into account the full range of uncertainties
in HD~209458's temperature, luminosity, and metallicity in the unpolluted scenario.
Given the size of the star's convection zone, pollution by $\sim40$~M$_{\oplus}$ of
metal material is required to produce the presumed metallicity enhancement of
$\Delta$Z=0.008.  If this were the case, orbital inclination estimates for the planet
HD~209458b would need to be revised to match the lower mass of its parent star.  
Following the $\chi^{2}$ lightcurve fit procedures described in CS02, the resulting
most probable planetary radius would be unaltered, but the mass uncertainty would
actually decrease by $\sim5$\%. However, the fit of the predicted lightcurve produced
by a 1.42~M$_{\rm J}$ planet to the observed one is not as good for a 0.97~M$_{\odot}$
parent star (the reduced $\chi^{2}$ value increases from 2.5 to 2.9).

For 100\% iron pollution, on the other hand, the best-fit polluted and unpolluted stellar
evolution tracks are virtually indistinguishable. This is because only
$\sim4$~M$_{\oplus}$ of pure iron pollution is required to raise the metallicity from
[Fe/H]=-0.2 to [Fe/H]=0.0. The resulting $\Delta$Z of 0.001 is not large enough to
substantially alter the stellar parameters.

\section{Discussion}

\begin{figure}[!h]
\begin{center}
\plotone{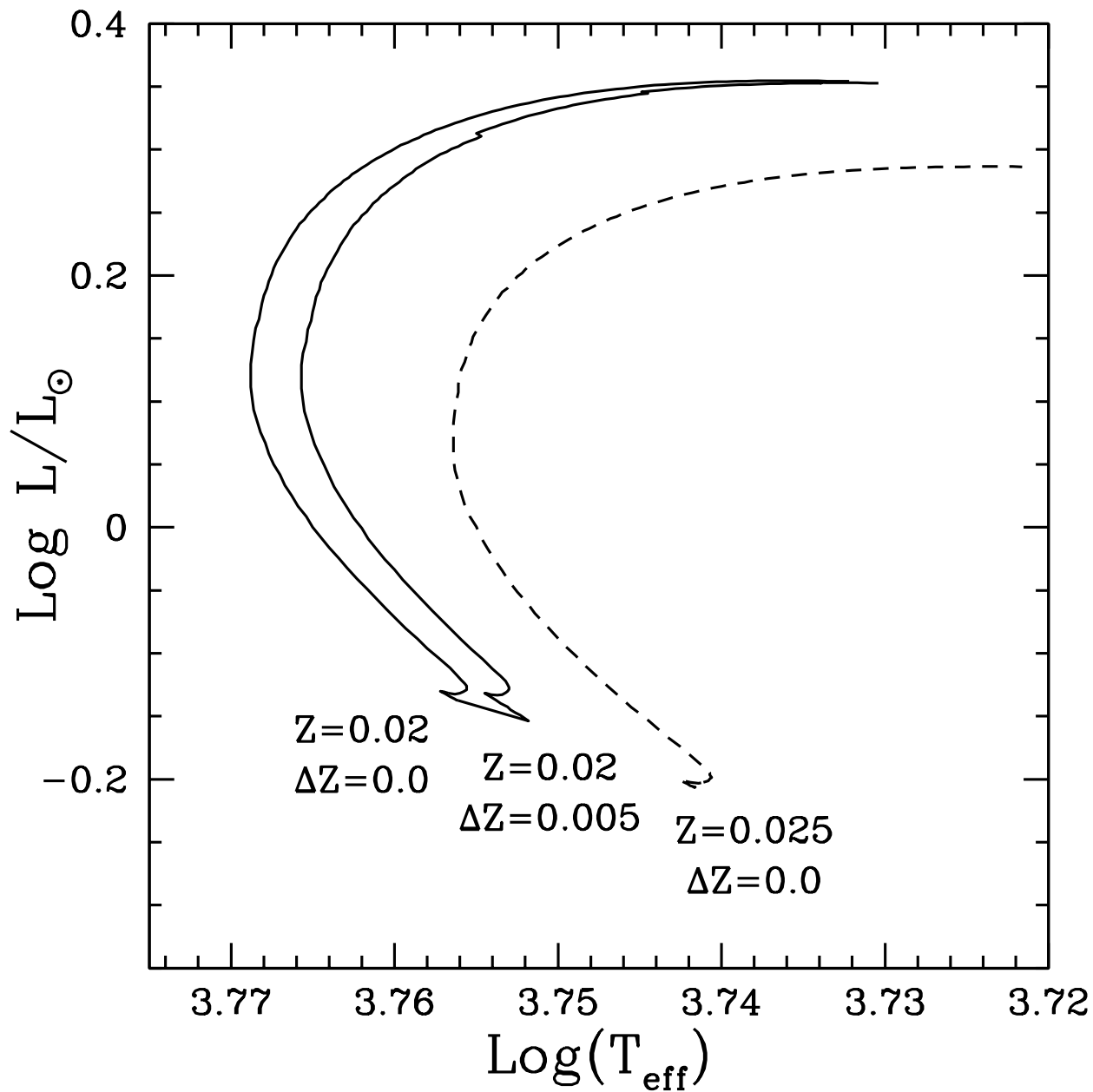}
\end{center}
\figcaption[1.0~M$_{\odot}$ Polluted Models]{Three 1.0~M$_{\odot}$ tracks with
(Z=0.02, $\Delta$Z=0), (Z=0.02, $\Delta$Z=0.005), and (Z=0.025, $\Delta$Z=0)
illustrate the relation between models containing uniform Z of 0.02, uniform
Z of 0.025, and interior Z of 0.02/surface Z of 0.025. Models
with matching interior metallicities are much closer to each other than those with
matching surface metallicities. \label{solarmass}}
\end{figure}

Our numerical calculations have confirmed the theoretical expectation that
pollution is capable of causing significant changes in the evolution of
stars.  We have shown that pollution from a primordial metal fraction
Z$_{\circ}$ to a final value Z$_{\circ}$+$\Delta$Z shifts evolution toward
lower temperatures, but the resulting evolutionary tracks are not coincident
with models of uniform metal fraction Z$_{\circ}$+$\Delta$Z. In fact, the
polluted tracks lie much closer on the H-R diagram to the Z=Z$_{\circ}$
models than those with Z=Z$_{\circ}$+$\Delta$Z throughout the entire star. In
Fig.\ \ref{solarmass}, we illustrate this effect for three stars of mass 1.0~M$_{\odot}$.  
The upshot of this result is that if a star of a particular mass is suspected
of having accreted a planet, its position on the H-R diagram should {\em not}
be consistent with the observed metallicity.  However, because accurate
stellar masses are rarely available, this condition may not be such a good
discriminant for the pollution scenario. Nevertheless, it could be useful in
cases where additional constraints are available, such as stellar clusters,
where isochrones fix the mass for a given temperature and luminosity.

Having explored the correlation between M$_{\rm cz}$, amount of metal enhancement, and
stellar temperature, we can address the prospects of directly observing abundance changes
due to pollution, and the quantities of material required to bring these enhancements
about. A number of authors have quoted figures on the order of a few M$_{\oplus}$ in their
estimates of pollution mass
\citep{2001ApJ...555..801M,2002ApJ...566..442M,2001ApJ...556L..59P}. But from the example
of HD~209458, it is evident that the amount of mass required is highly dependent on the
iron content of the material. If pure metal pollution with iron and other species in {\em
solar proportions} is added to a convection zone of mass M$_{\rm cz}$, the total amount of
pollution needed to raise the metal fraction from Z to $\Delta$Z is given by M$_{\rm
poll}$=M$_{\rm cz}\Delta$Z/(1-Z-$\Delta$Z).  Thus, for a 1.15~M$_{\odot}$ star with
M$_{\rm cz}\sim$0.007~M$_{\odot}$ (as shown in Fig.\ \ref{conmass}), and pollution from
Z$_{\odot}$=0.02 to 0.03, we find M$_{\rm poll}\sim$25~M$_{\oplus}$.  And as seen in the
hypothetical scenario for HD~209458 (M$_{\rm poll}\sim$40~M$_{\oplus}$), the necessary
M$_{\rm poll}$ would be somewhat larger. On the other hand, if the metal pollutant is {\em
more} iron rich than solar material, as little as one tenth of this mass would be needed 
to achieve the same surface metallicity enhancements.

The question of what sort of planetary bodies are most likely to serve as
contributors to pollution is even more uncertain. Recent theories have proposed that
gas giant planets with rocky cores may experience orbital migration and possibly fall
into their host stars. But the amount of metal available in the cores of these
Jupiter-like objects is not well known, even for members of our own solar system.
Models developed by \citep{1997Icar..130..534G} and \citep{1999Sci...286...72G}
estimate that Jupiter may possess a rocky core of $<12$M$_{\oplus}$, and it contains
a total heavy-element mass between 11 and 45 M$_{\oplus}$. Yet, even if a planet with
a metallic core of $12{\rm M}_{\oplus}$ were to pollute its host star, the gaseous
atmosphere would also supply an amount of hydrogen and helium, thereby diluting the
contribution of high-Z material. In the case of Jupiter the average Z is between
$\sim$3 and 13\%, and for Saturn, it is $\sim$20-30\% \citep{1999P&SS...47.1175H}.  
Hence, due to the $\gtrsim$70\% contribution of hydrogen and helium, the mass of
accreted planetary matter involved in pollution will inevitably be greater than the
values quoted in our pure metal enrichment models. One way around the issue of
limited metal content in gas giant planets is ingestion of smaller planetesimals.
Because most planets systems are thought to have formed in a circumstellar dust disk,
there may be enough rocky material available to pollute stars during the first 10
million years of their lifetimes, before the disk dissipates
\citep{1997MNRAS.285..403G}.  Accretion of planetesimals certainly seems feasible,
but the large convection zone depths in very young stars poses problems regarding the
observability of such an effect \citep{2001ApJ...555..801M}. Hence, at present, there
does not appear to be a universally favored method for large quantities of
heavy-element material to accrete onto stars and remain reasonably concentrated at
the stellar surface.

\begin{figure}[!h]
\begin{center}
\plotone{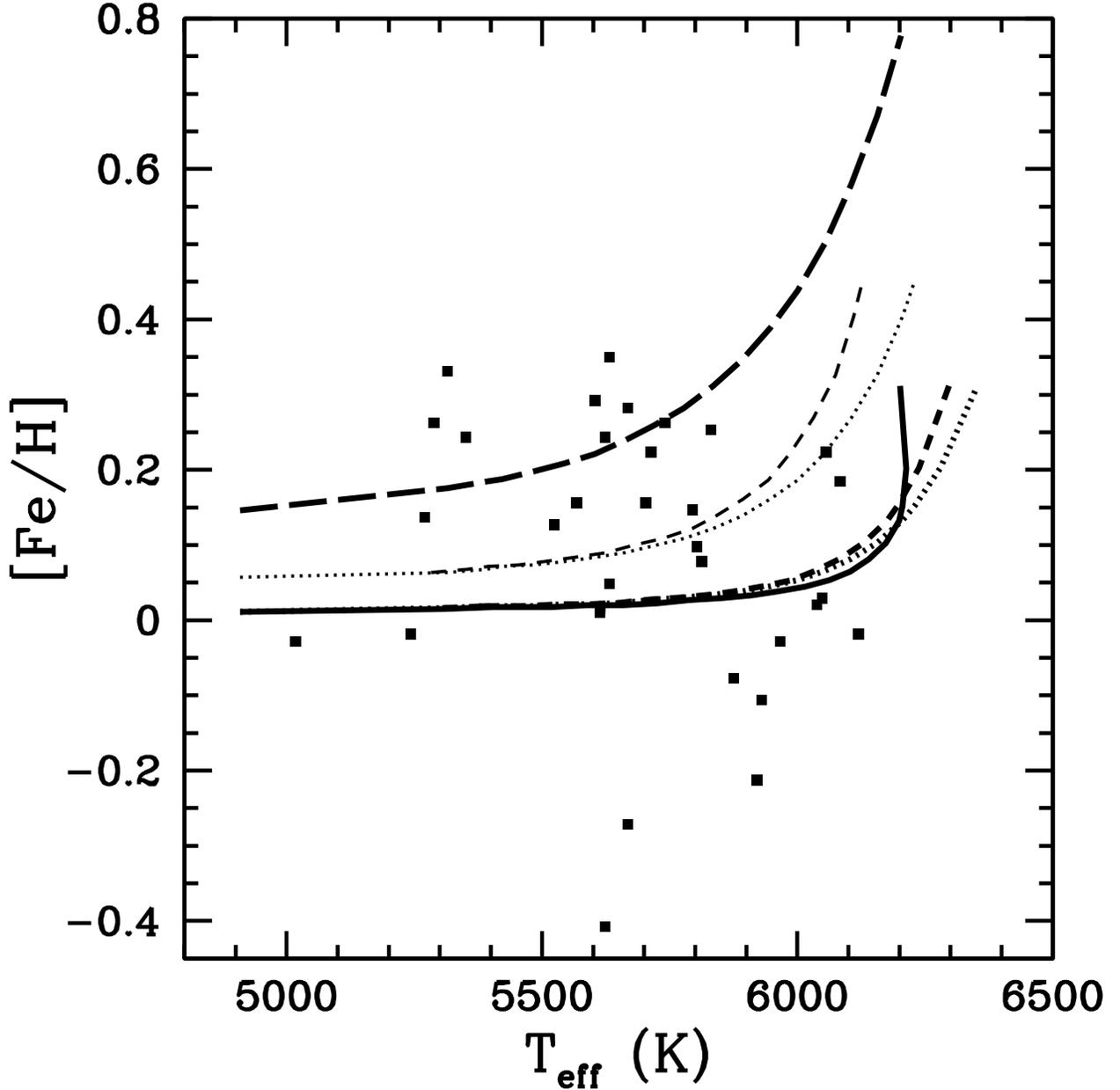}
\end{center}
\figcaption[Fixed pollution versus mass]{Amount of surface pollution as a function
of stellar effective temperature. The latter correlates strongly with stellar
mass (and hence - convection zone mass) on the main sequence. The points are
nearby stars with known giant planets. The thin curves show metallicities expected for
stars that accrete 40~M$_{\oplus}$ of pure metals (in solar proportions) on the ZAMS, 
while three lower thick curves depict metallicities calculated for stars that accrete 
10~M$_{\oplus}$ of metal (also in solar proportionas). We have also included one 
model with 10~M$_{\oplus}$ of pure iron accretion (topmost curve). All models start 
with Z$_{\circ}$=0.02. We have plotted the metallicity trends for several different ages 
during the subsequent polluted evolution: 0.1~Myr (dotted lines and long dashes), 
0.3~Myr (short dashes), and 3.0~Gyr (solid). Our models suppress the large enhancement
at high temperatures seen in previous studies, but are still a poor fit to the 
observations. 
\label{fixed}}
\end{figure}

Finally, we examine the viability of pollution in light of current observational data
on samples of SWPs. Exploring model behavior for a range of masses and fixed M$_{\rm
poll}$, rather than fixed $\Delta$Z (as in Fig.\ \ref{zconst}) enables us to predict
what trends, if any, should be present among the observations if pollution is indeed
a common process. We have polluted solar-metallicity (Z$_{\circ}$=0.02) stars on the
ZAMS with 10~M$_{\oplus}$ and 40~M$_{\oplus}$, and subsequently evolved them to
3~Gyr. Fig.\ \ref{fixed} depicts the temperatures and surface metallicities expected
at several different ages. This is similar to Fig.~2 in \citet{2001ApJ...556L..59P},
but our models include {\em polluted} evolution. Stars with higher temperatures, and
hence larger masses, have smaller convection zones; thus is it no surprise that we
find significant increases in surface metallicity for the high-temperature models.
For comparison, we have overplotted data points from a recent study by
\citet{2003scifro..6..392F}. Pollution would undoubtedly contribute some range of
M$_{\rm poll}$ to stars, as opposed to the fixed quantities 10~M$_{\oplus}$ and
40~M$_{\oplus}$ that we have selected.  We have also simplified the situation 
by only showing stellar models that have initial metal compositions of
Z$_{\circ}$=Z$_{\odot}$=0.02. But a superposition of [Fe/H] versus T$_{\rm eff}$ 
models covering a {\em range} of initial metallicities and pollution masses should still 
produce preferential surface metallicity enhancement among the high-temperature data points 
if planet accretion is widespread. No such trend is evident. 
The presence of older, cooler stars in the data could also skew the results, since the 
convection zones of evolved stars are deeper, and hence the effects of pollution would be 
diluted. But the sample here includes only five subgiants, and \citet{2003scifro..6..392F} 
note that none of them have particularly low metallicities. Thus the current data suggest 
that pollution cannot be the only cause behind the relatively high metallicities of 
stars with planets.

\section{Conclusions}
In modifying the Princeton Stellar Evolution code, we have paved the way for detailed
study of stellar structure and evolution as affected by planet accretion. Grids
of polluted models for various masses and metallicities have enabled us to identify the
main physical outcomes of this process. The most prominent features are:
\begin{itemize}
\item Deepening of the stellar convection zone 
\item Expansion of the star, leading to lower effective temperatures at approximately
constant luminosity
\item Age-independence of the results
\end{itemize}

  We expect that the general trends presented here will be useful in future attempts to
discriminate between pollution and other sources of metal enhancement in stars with
planets. In addition, they should be particularly well suited to studies of pollution
binary systems or star clusters in which additional parameters constrain stellar age
and mass.  Pollution models applied to specific candidates in which there is
information on the stellar mass or age could also offer fruitful methods of
determining whether planet consumption was part of a star's past. In
subsequent work, we develop polluted models of several SWPs. We will present
predictions for p-mode oscillation frequencies in these systems, with the goal of
directly testing for pollution with current asteroseismology missions.

\acknowledgements{AMC is grateful for support from the Harvard College
Research Program throughout this work. We acknowledge additional support from
the Harvard Clark Fund. Sincere thanks also go to Douglas Gough and to the 
referee for useful feedback.}

\clearpage
\bibliographystyle{apj}
\bibliography{pollution}

\clearpage

\end{document}